\documentclass[floatfix,groupedaddress,superscriptaddress,aps,showpacs,amsmath,amssymb,floatfix, prx, longbibliography, twocolumn,a4]{revtex4-2}

\usepackage{graphicx,float}% Include figure files
\usepackage{subfigure}
\usepackage{dcolumn}% Align table columns on decimal point
\usepackage{bm}% bold math
\usepackage{mathrsfs}
\usepackage{txfonts}
\usepackage{CJK}
\usepackage[amsmath]{SIunits}
\usepackage{epsfig}
\usepackage{epstopdf}
\usepackage{lipsum}
\usepackage{color}
\usepackage{placeins}
\usepackage{lipsum}

\usepackage[colorlinks,
linkcolor=blue,
anchorcolor=blue,
citecolor=blue,
filecolor=blue,
menucolor=blue,
runcolor=blue,
urlcolor=blue,
frenchlinks=blue]{hyperref}

\newcommand{\ket}[1]{|#1\rangle}
\newcommand{\bra}[1]{\langle #1|}
\allowdisplaybreaks[2]
\newenvironment{SChinese}{%
\CJKfamily{gbsn}%
\CJKtilde
\CJKnospace}{}
\allowdisplaybreaks[2]

\begin{document}

\begin{CJK}{UTF8}{}
\begin{SChinese}

\title{Nonreciprocal $\mathcal{PT}$-symmetric phase transition in a non-Hermitian chiral quantum optical system}

\author{Miao Cai}  %
\thanks{These two authors contributed equally}
 \affiliation{College of Engineering and Applied Sciences, National Laboratory of Solid State Microstructures, and Collaborative Innovation Center of Advanced Microstructures, Nanjing University, Nanjing 210023, China}

\author{Jiang-Shan Tang}  %
\thanks{These two authors contributed equally}
 \affiliation{College of Engineering and Applied Sciences, National Laboratory of Solid State Microstructures, and Collaborative Innovation Center of Advanced Microstructures, Nanjing University, Nanjing 210023, China}
 \affiliation{Hefei National Laboratory, Hefei 230088, China}
 
 \author{Ming-Yuan Chen}  %
 \affiliation{College of Engineering and Applied Sciences, National Laboratory of Solid State Microstructures, and Collaborative Innovation Center of Advanced Microstructures, Nanjing University, Nanjing 210023, China}
 
\author{Keyu Xia}  %
 \email{keyu.xia@nju.edu.cn}
    \affiliation{College of Engineering and Applied Sciences, National Laboratory of Solid State Microstructures, and Collaborative Innovation Center of Advanced Microstructures, Nanjing University, Nanjing 210023, China}
    \affiliation{Hefei National Laboratory, Hefei 230088, China}

\date{\today}

\begin{abstract}
Phase transitions, non-Hermiticity and nonreciprocity play central roles in fundamental physics. However, the triple interplay of these three fields is of lack in the quantum domain. Here, we show nonreciprocal parity-time-symmetric phase transition in a non-Hermitian chiral quantum electrodynamical system, caused by the directional system dissipation. In remarkable contrast to previously reported nonreciprocal phase transitions, the nonreciprocal parity-time-symmetric phases appear even when the atom-resonator coupling is reciprocal. Nonreciprocal photon blockade is obtained in the nonreciprocal phase region. These results may deepen the fundamental insight of nonreciprocal and non-Hermitian quantum physics, and also open a new door for unconventional quantum manipulation. 
\end{abstract}

\maketitle

\end{SChinese}
\end{CJK}

%Introduction
\emph{Introduction}.---Phase transitions, characterized by uncontinuous change of system properties driven by external parameters, are crucially important to fundamental physics. Phenomena related to phase transitions range from our daily-life experience of ice melting to far more exotic superfluid Mott-insulator phase transitions~\cite{Greiner2002}, from microscopic quark-level matter transitions~\cite{Chapline1976} to cosmological phase transitions in the early Universe~\cite{Kibble1980}, and from single-particle~\cite{Wang1973} to many-body Dicke quantum phase transitions~\cite{Baumann2010}. The investigation on zero-temperature phase transitions, for the first time, reveals the role of quantum effects in phase transitions, opening up the field of quantum phase transition~\cite{Kravchenko1995, Bitko1996, Subir2000, Luo2017}. Quantum phase transitions are vital for understanding the evolution of matter phases~\cite{Sondhi1997, Subir1999, Matthias2003} and have been widely observed or predicted in various systems ranging from condensed-matter systems~\cite{Orenstein2000} and atomic systems~\cite{Greiner2002} to optical systems~\cite{Greentree2006, Mann2018, Wang2024}. 

A non-Hermitian system can display simultaneous phase transitions associated with the parity-time ($\mathcal{PT}$) symmetry breaking at its exceptional point (EP)~\cite{Bender2007}. A variety of exotic physics have been revealed in a non-Hermitian system related to the $\mathcal{PT}$-symmetric phase transitions~\cite{Cejnar2007, Lee2014, Zhou2018, Longhi2019, Liu2019, Matsumoto2020, Dai2024}. 
A quantum single spin can also exhibit the $\mathcal{PT}$-symmetric phase transition~\cite{Peng2016, Chao2023, Wu2024, Ning2024}. Now the non-Hermitian system emerges as a highly appealing platform for studying phase transitions in both the classical and quantum domains.

Nonreciprocity is a concept equally important to phase transition and non-Hermiticity.  
It plays an indispensable role in many important applications such as optical isolation~\cite{Yu2009, Reed2010, Lira2012, Estep2014, Xia2014, Sayrin2015, Michael2016, Ruesink2016, Zhang2018, XiaPRL2018, Huang2018, Yang2019, XiaSciAdv2021, Hu2021, TangPRL2022, TangLei2022, Pan2022} and sensing enhancement~\cite{Fleury2015, Lau2018, Li2019, McDonald2020, Jiao2020, Bao2021, Hashemi2022,Ding2023, Mao2024}. 
Nonreciprocity shows parallel importance in fundamental physics. Nonreciprocal interaction between different parties of a system, as a vital mechanism to build a non-Hermitian system, has been explored to study one-way topological solitons in active matter~\cite{Veenstra2024}, novel phase transitions~\cite{Kryuchkov2018, Fruchart2021, Zhang2021, Banerjee2022, Chiacchio2023, Dinelli2023, Alston2023, Verma2024}, the skin effect~\cite{Tang2023, Aoxi2023}, band structure in a Floquet medium~\cite{Jagang2022}. Nonreciprocity is even used to distinction of fundamental quantum effects~\cite{Wu2022}. By breaking the time-reversal ($\mathcal{T}$) symmetry via nonreciprocal modulation, a non-Hermitian Aharonov-Bohm effect and phase transitions with unconventional phononic $\mathcal{PT}$ symmetry are observed in an optomechanical resonator~\cite{Dai2024}. Despite great advances, nonreciprocal phase transitions in a non-Hermitian quantum system remains unexplored so far in the case without intermode nonreciprocal interaction or control.

In this work, by leveraging chiral dissipation induced by breaking the $\mathcal{T}$ symmetry of a non-Hermitian cavity quantum electrodynamic system (cQED), we show nonreciprocal $\mathcal{PT}$-symmetric phase transitions \emph{without the need of nonreciprocal interaction or modulation}. In combination with a chiral atom-resonator coupling, we attain reciprocal photon transmission but strong nonreciprocal quantum correlation, implying nonreciprocal photon blockade.

\begin{figure}
  \centering
  \includegraphics[width=1.0\linewidth]{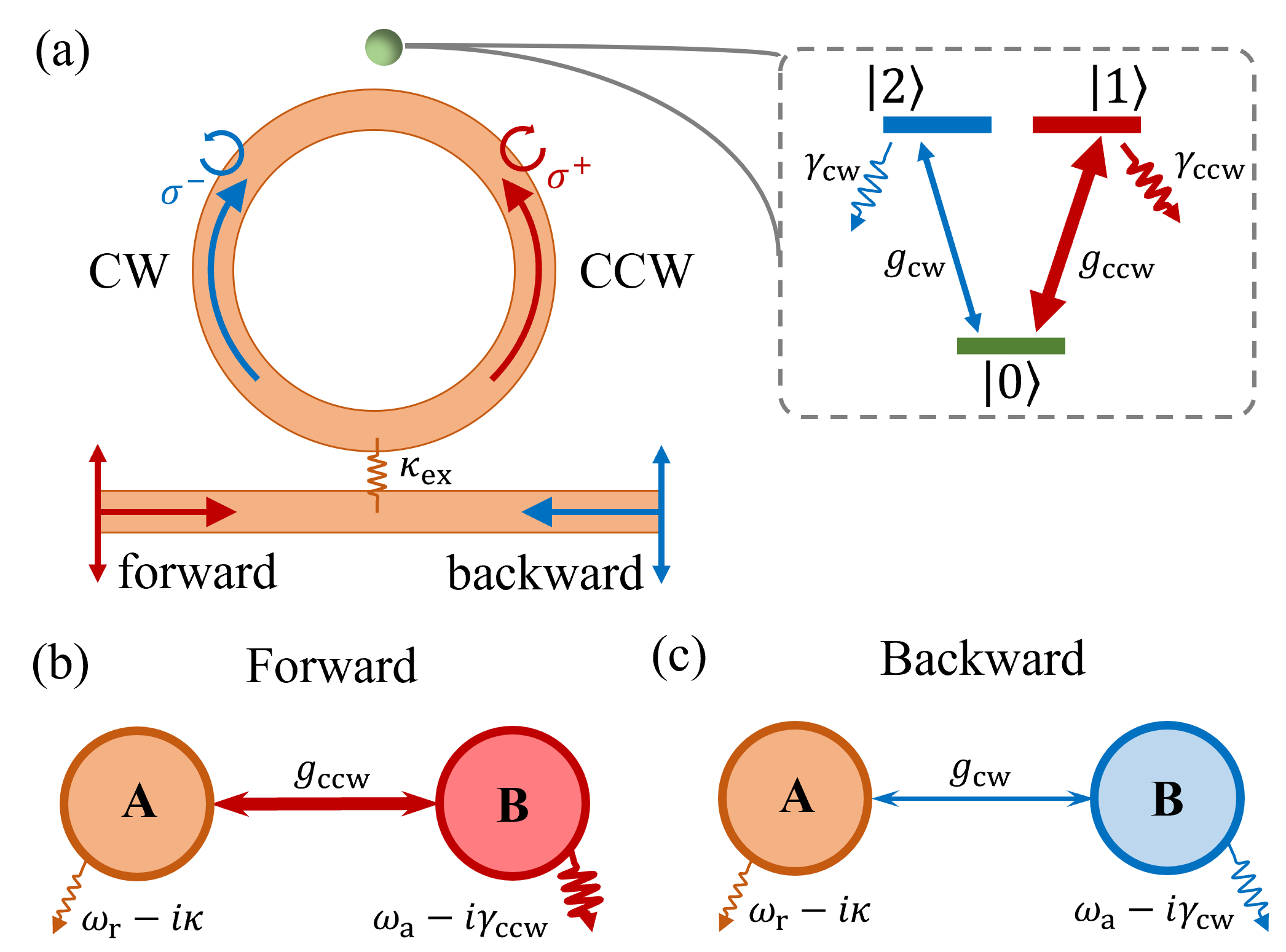} \\
\caption{(a) Schematic for nonreciprocal $\mathcal{PT}$-symmetric phase transition in a non-Hermitian chiral cQED system. The CCW (CW) mode couples to the V-type atom via its $\sigma_+$-polarized ($\sigma_-$-polarized) evanescent field and drives the atomic transition $\ket{0} \leftrightarrow \ket{1}$ ($\ket{0} \leftrightarrow \ket{2}$) with strength $g_\text{CCW}$ ($g_\text{CW}$). The states $\ket{1}$ and $\ket{2}$ respectively decay with rates $\gamma_\text{CCW}$ and $\gamma_\text{CW}$. An optical waveguide is used to probe the nonreciprocal behaviors of the system with the forward and backward inputs. The CW and CCW modes decays with rate $\kappa_\text{in}$ intrinsically and with rate $\kappa_\text{ex}$ externally to the waveguide.  (b) and (c) Single-excitation non-Hermitian model for the forward and backward cases.}
\label{fig:FIG1}
\end{figure}

%
% Model system 1
\emph{System and model}.---The non-Hermitian chiral cQED system is schematically shown in Fig.~\ref{fig:FIG1}(a). It consists of a $V$-type atom coupling to a microring resonator. To probe the system dynamics, a single-mode optical waveguide is coupled to the resonator. This waveguide causes the fields in the resonator to decay with rate $\kappa_\text{ex}$. 

The resonator supports two whispering-gallery modes (WGMs) propagating in the counterclockwise (CCW) and clockwise (CW) directions, respectively. We dub the system dynamics associated with the CCW and CW modes as the forward and backward cases or the CCW and CW cases accordingly. These CW and CCW modes are degenerate in resonance frequency, denoted as $\omega_r$, and intrinsically decay with rate $\kappa_\text{in}$, yielding the same overall decay rate $\kappa = \kappa_\text{in} + \kappa_\text{ex}$. Noticeably, the evanescent fields of their quantum vacuum fluctuation can be circularly polarized near the resonator surface. Their polarizations are directional under time reversal and thus chiral~\cite{Lodahl2017, Tang2019}. According to numerical simulations~\cite{Tang2019} and theoretical analysis~\cite{Mechelen2016}, the evanescent field of the CCW (CW) mode is left-hand (right-hand) circularly polarized, denoted as $\sigma_+$ ($\sigma_-$). 
%This resonator can be a bottle-shaped fiber~\cite{ArnoPRX, ArnoScience} or a loop optical waveguide integrated on a chip~\cite{XiaPRA2019}.

The quantum vacuum field of the resonator mode can strongly interacts with an atom nearby the resonator surface. We consider a V-type atom coupling to the resonator. The ground and two excited states of the atom are $\ket{0},\ket{1},\ket{2}$. The CCW and CW modes couples to the transitions $\ket{0}\leftrightarrow\ket{1}$ and $\ket{0}\leftrightarrow\ket{2}$, with strengths $g_\text{CCW}$ and $g_\text{CW}$, respectively. The states $\ket{1}$ and $\ket{2}$ decay with rates $\gamma_\text{CCW}$ and $\gamma_\text{CW}$. We consider the case of $\gamma_\text{CCW} > \gamma_\text{CW}$, $\kappa \neq \gamma_\text{CW}$ and $\kappa \neq \gamma_\text{CCW}$, leading to non-Hermiticity in both the forward and backward cases. The system Hamiltonian including the atom-resonator interaction and the dissipation is given by $H_s/\hbar = (\omega_{\text{a}} -i \gamma_\text{CCW})\sigma_{11} + (\omega_{\text{a}} -i \gamma_\text{CW})\sigma_{22} + (\omega_{\text{r}} - i\kappa) (a_\text{CW}^\dag a_\text{CW} + a_\text{CCW}^\dag a_\text{CCW}) + g_{\text{cw}}(a_\text{CW}^\dag \sigma_{02} + \sigma_{20}a_\text{CW}) + g_{\text{ccw}}(a_\text{CCW}^\dag \sigma_{01} + \sigma_{10}a_\text{CCW})$, where the atomic operators are defined as $\sigma_{fi} = \ket{f}\bra{i}$, $a_\text{CW}$ and $a_\text{CCW}$ are the annihilation operators of the CW and the CCW mode, respectively, $\hbar$ is the Planck constant. Here, we set the ground state as the reference level. To probe the nonreciprocal behavior of the system, we input weak fields $\alpha_\text{in}$ and $\beta_\text{in}$ with frequency $\omega_p$ in the forward and backward directions, respectively. The Hamiltonian describing this probe takes the form $H_D / \hbar = i\sqrt{2\kappa_\text{ex}} \alpha_\text{in} a_\text{CW}^\dag + i \sqrt{2\kappa_\text{ex}}\beta_\text{in} a_\text{CCW}^\dag + H.c.$ with  $|\alpha_\text{in}|^2$ and $|\beta_\text{in}|^2$ representing the input photon fluxes. Then, the system dynamics is determined by the quantum master equation,
\begin{equation}
\label{eq:Eq1}
\begin{aligned}
	\dot{\rho} = & -i \left[H/\hbar, \rho\right] + 2 \kappa a_\text{CW} \rho a_\text{CW}^\dag +2 \kappa a_\text{CCW} \rho a_\text{CCW}^\dag \\
	& + 2 \gamma_{\text{CCW}}  \sigma_{01} \rho  \sigma_{10} + 2 \gamma_{\text{CW}}  \sigma_{02}  \rho  \sigma_{20} \;,
\end{aligned}
\end{equation}
where $H= H_s + H_D$, $\rho$ is the system density matrix. The CCW and CW modes are orthogonal in polarization and drive the two atomic transitions separately. Thus, the system exhibits chiral atomic dissipation under time reversal. To characterizing the transmission and full-quantum dynamics of the system, we numerically solve the master equation and truncate the resonator mode to the fourth photonic Fock state.

The key point of this work is that the cQED system is non-Hermitian and possesses quantum chirality, induced by breaking the $\mathcal{T}$ symmetry. Below, we show these properties and the system eigenfrequencies in the single-excitation space. Under weak-input condition, the system reduces to two decoupled cQED subsystems. Each includes a resonator mode and a two-level atom but with different dissipation rates, thus is non-Hermitian. Particularly, in the single-excitation space, these non-Hermitian subsystems can be modeled as two-mode systems consisting of a cavity mode A (orange circle) coupling to a cavity mode B (red or blue circle) with strength $g_\text{CCW}$ and $g_\text{CW}$, as shown in Figs.~\ref{fig:FIG1}(b) and (c). The mode A maintains the loss rate of $\kappa$, whereas the mode B decays with rate $\gamma_\text{CCW}$ and $\gamma_\text{CW}$ in the forward and backward cases, respectively. 
Thus, the $\mathcal{T}$ symmetry breaks and we ``see'' two different non-Hermitian subsystems in opposite directions. The Hamiltonians for these two subsystems including the mode dissipation can be written as~\cite{Choi2010},
\begin{equation}
\label{eq:Eq2}
	H_\text{x} = \left(\begin{array}{ccc}
	 \omega_{\text{r}} - i\kappa  & g_\text{x} \\
	g_\text{x} & \omega_{\text{a}} - i\gamma_\text{x}
	\end{array}\right) \;,
\end{equation}
with $\text{x} = \text{``CW''}$ and $\text{x} = \text{``CCW''}$ representing the CW and CCW cases, respectively. Obviously, $H_\text{CCW} \neq H_\text{CW}$ when $\gamma_\text{CCW} \neq \gamma_\text{CW}$ or $g_\text{CCW} \neq g_\text{CW}$ or both, displaying nonreciprocity. The eigenfrequencies of this Hamiltonian are $E_{\pm,\text{x}}  = (\omega_{+} - i\gamma_{+,\text{x}}) \pm \sqrt{(\omega_{-} - i\gamma_{-,\text{x}})^2 + g^2}$,
where $\omega_{\pm} = (\omega_{\text{r}} \pm \omega_{\text{a}})/2$ and $\gamma_{\pm,\text{x}} = |\kappa \pm \gamma_{\text{x}}|/2$. Under the atom-resonator resonance condition that $\omega_\text{a} = \omega_\text{r} = \omega_0$, yielding $\omega_- =0$ and $\omega_+ = \omega_0$, the eigenvalues become 
\begin{equation}
\label{eq:Eq3}
	E_{\pm,\text{x}}  = (\omega_0 - i\gamma_{+,\text{x}}) \pm \sqrt{g^2 - \gamma_{-,\text{x}}^2} \;.
\end{equation}
We consider $\gamma_{-,\text{x}} \neq 0$ in both the CW and CCW cases. Thus, the two subsystems are non-Hermtinian. Both the eigenvectors and eigenfrequencies coalesce at EP $g_{\text{EP},\text{x}}= \gamma_{-,\text{x}}$, associated with the $\mathcal{PT}$-symmetric phase transition. Because $g_{\text{EP}, CW} \neq g_{\text{EP}, CCW}$, phase transitions can occur at different coupling strength in two cases. Thus, the original cQED system can display nonreciprocal $\mathcal{PT}$-symmetric phase transitions in the quantum domain when $g_{\text{EP}, CW} < g_\text{CW} < g_{\text{EP}, CCW} < g_\text{CCW}$ as the distance $r$ between the atom and the resonator surface varies. We define the detuning between the input probe field and the atom-resonator system as $\Delta_{\text{p}} = \omega_{0} - \omega_{\text{p}}$.

A rubidium or cesium (Cs) atom nearby a bottle-shaped or integrated microring resonator can be used to realize our chiral cQED system~\cite{Xia2014, Sayrin2015, Michael2016, Tang2019}. Throughout our investigation below, we consider a Cs atom resonantly coupling to the resonator via the evanescent quantum vacuum fields of the CW and CCW modes such that $\omega_- = 0$ and $\omega_+ = \omega_0$. 
 The Cs atom is prepared in the ground state $\ket{6^2S_{1/2}, F=4, m_F = 4}$. The two excited states are $\ket{1} = \ket{6^2P_{3/2}, F^\prime=5, m^\prime_F = 5}$ and $\ket{2} = \ket{6^2P_{3/2}, F^\prime=5, m^\prime_F = 3}$. The atom couples to the $\sigma_+$- and $\sigma_-$-polarized fields with different dipole moments $d_+$ and $d_-$ that $d_+ = \sqrt{45} d_-$~\cite{Daniel2007}. Thus, we have $g_{\text{CW}} = g_{\text{CCW}}/\sqrt{45}$ and $\gamma_{\text{CW}} = \gamma_{\text{CCW}}/45$. Note that the coupling strength exponentially increases as the atom-surface distance $r$ decreases. This allows us to tune the atom-resonator coupling and dynamically study the $\mathcal{PT}$-symmetric phase transition of the system.
 For simplicity, we choose the following normalized system parameters for our model and numerical simulations: $\gamma_\text{CW} = 1, \gamma_\text{CCW} = 45$, $\kappa_\text{in} = 0.5$, $\kappa_\text{ex} = 2.5$ that $\gamma_\text{CW} < \kappa_\text{ex} < \gamma_\text{CCW}$, yielding $g_\text{EP, CCW} = 21$ and $g_\text{EP, CW} = 1$.  

\begin{figure}
  \centering
  \includegraphics[width=1.0\linewidth]{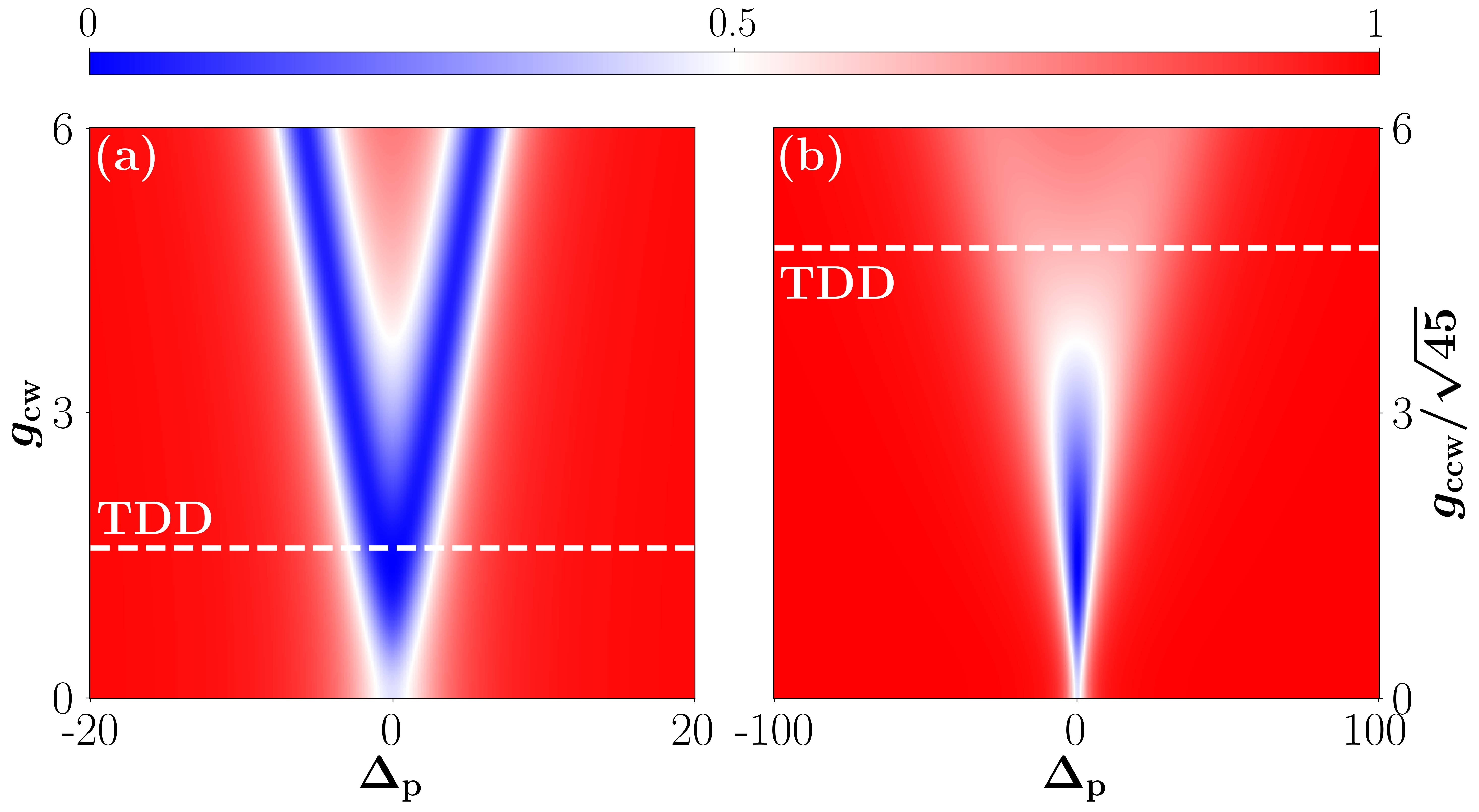} \\
\caption{Resonator transmission with respect to the atom-resonator coupling strength and the frequency detuning under (a) the CW mode and (b) the CCW mode excitation. The white dashed lines labeled with ``TDD'' indicate the critical coupling strengths where transmission dip degeneration happens.}
\label{fig:FIG2}
\end{figure}

\emph{Nonreciprocal transmission}.---For the purpose of experimental test, we use the weak-probe transmission to characterize the nonreciprocal $\mathcal{PT}$-symmetric phase transition in our chiral cQED system. To do so, we firstly investigate the nonreciprocal resonator transmission in opposite directions.  The resonator transmission versus the atom-resonator coupling strength and the probe detuning $\Delta_p$ are presented in Fig.~\ref{fig:FIG2}. In comparison of the transmissions, the relation $g_\text{CW} = g_\text{CCW}/\sqrt{45}$ maintains. Thus, we plot figures by scalling the CCW coupling strength as $g_{\text{CCW}}/\sqrt{45}$. The transmission spectra gradually change from singlet-dip to doublet-dip profiles as the coupling strength increases in both cases. By setting the derivative of the transmission zero~\cite{TangTPD2022}, we obtain the points of transmission dip degeneracy (TDD), corresponding to a critical coupling strength
\begin{equation}
\label{eq:Eq4}
	g_{\text{TDD,x}} = \sqrt{\left(-\xi_{\text{TDD},x} + \sqrt{\xi_{\text{TDD},x}^2 - 4\zeta_{\text{TDD},x}\eta_{\text{TDD},x}}\right)/2\zeta_{\text{TDD},x}}  \;,
\end{equation}
where $\text{x} \in \{\text{CW, CCW}\}$, $\zeta_\text{TDD,x} = 2\gamma_\text{x} + \kappa_{\text{in}}$, $\xi_\text{TDD,x} = 2\gamma_\text{x}^2\kappa_{\text{in}} - \gamma_\text{x}^3 - \gamma_\text{x}\kappa_\text{ex}^2 + \gamma_\text{x}\kappa_{\text{in}}^2$ and 
$\eta_\text{TDD, x} = -\kappa_\text{in}\gamma_\text{x}^4$. When the coupling exceeds these critical values, the transmission spectral splittings occur. Note that a non-Hermitian sensor exhibits unprecedented enhancement of sensitivity around the transmission peak degeneracy~\cite{Kononchuk2022}. The TDD may have the potential in sensing enhancement.

By substituting system parameters into Eq.~\ref{eq:Eq4}, we obtain $g_{\text{TDD,CW}} = 1.58$ and $g_{\text{TDD,CCW}} = 31.78 \approx 4.74\sqrt{45}$ in the CW and CCW case, respectively. The critical coupling strengths for TDD are marked with white dashed lines in Fig.~\ref{fig:FIG2}. As the atom-surface distance decreases, both the coupling strengths $g_\text{CW}$ and $g_\text{CCW}$ increase. Thus, $g_{\text{TDD,CW}} \neq g_{\text{TDD,CCW}}/\sqrt{45}$ implies the different critical atom-surface distance in two cases where the transmission spectrum splits. These weak-probe transmissions provide an experimentally accessible way to extract the system eigenfrequency in the single-excitation space.

\begin{figure}
  \centering
  \includegraphics[width=1.0\linewidth]{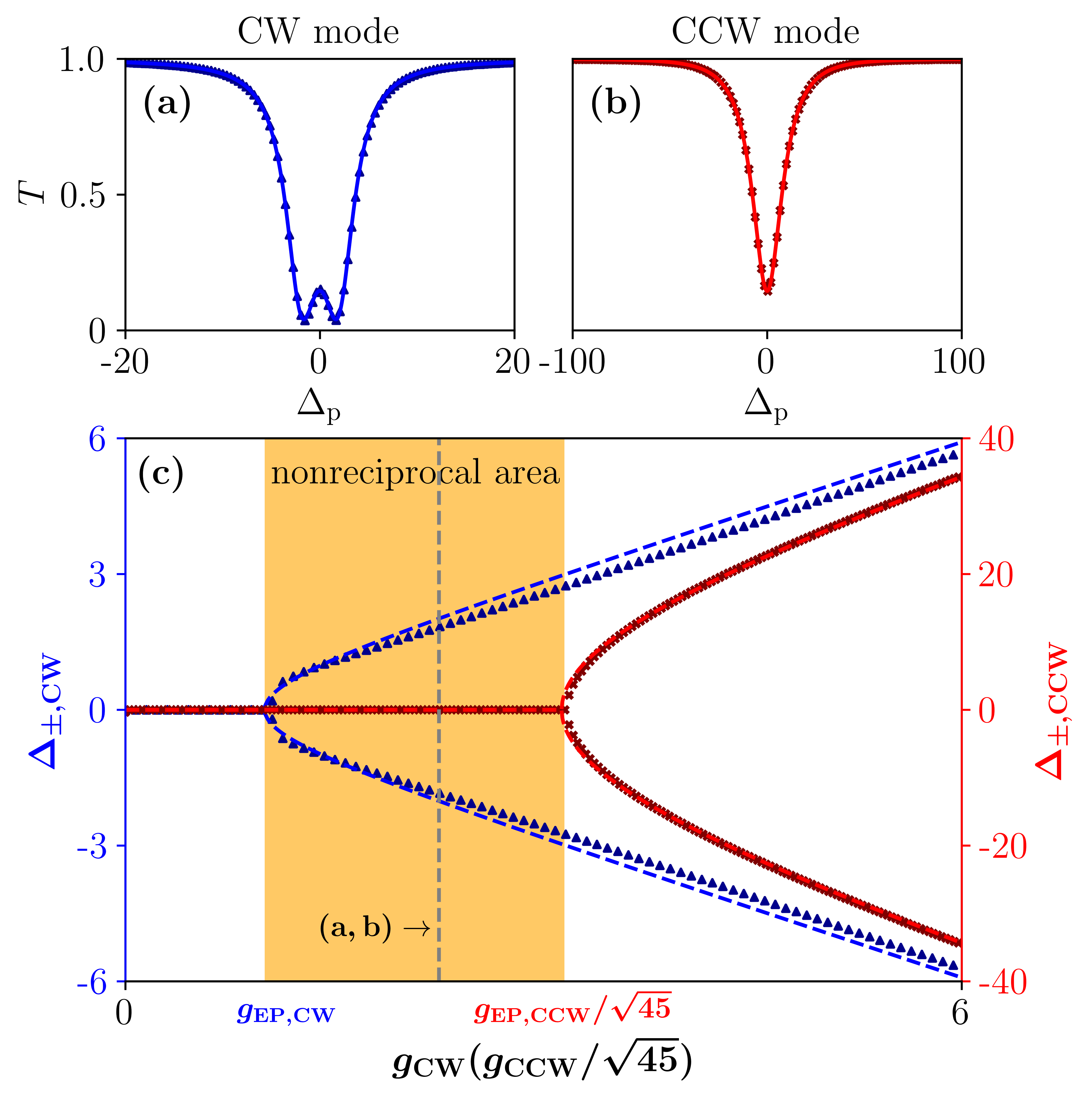} \\
\caption{(a) Transmissions at $g_\text{CW} = 2.25$ in the forward case. (b) Transmissions at $g_\text{CCW} = \sqrt{45}g_\text{CW} = 15.09$ in the backward case. In (a) and (b), solid curves are results of numerical simulation. Dots represent fitting using Eq.~\ref{eq:Eq5}. (c) Eigenfrequencies with respect to $\omega_p$ in the forward (red markers and curves) and backward (blue markers and curves) cases. The markers are numerical results. The dashed curves are fittings.}
\label{fig:FIG3}
\end{figure}

\emph{Nonreciprocal phase transition}.---Below we characterize the nonreciprocal $\mathcal{PT}$-symmetric phase transition with the resonator transmission. In our system, the eigenfrequencies can be extracted from the resonator transmission spectrum by fitting it with the following function~\cite{Kenta2021},
\begin{equation}
\begin{aligned}
\label{eq:Eq5}
	T(\Delta_\text{p}) &\sim  \\
	 &1- c_{1}\frac{\Gamma_{+,\text{x}}\Gamma_{-,\text{x}}}{ \left[(\Delta_\text{p} - \Delta_{+,\text{x}})^2 + \Gamma_{+,\text{x}}^2\right] \left[(\Delta_\text{p} - \Delta_{-,\text{x}})^2 + \Gamma_{-,\text{x}}^2\right]}\\
	&- c_{2} \frac{\Gamma_{+,\text{x}}^2\Gamma_{-,\text{x}}^2}{ \left[(\Delta_\text{p} - \Delta_{+,\text{x}})^2 + \Gamma_{+,\text{x}}^2\right]^2 \left[(\Delta_\text{p} - \Delta_{-,\text{x}})^2 + \Gamma_{-,\text{x}}^2\right]^2 } \;,\\
	& \text{x}\in\{\text{CW, CCW}\}\; ,
\end{aligned}
\end{equation}
where $\Delta_{\pm,\text{x}} = \text{Re}(E_{\pm,\text{x}}) - \omega_{0}$ are the detunings between the eigenmodes and the probe field, $\Gamma_{\pm,\text{x}} = \text{Im}(E_{\pm,\text{x}})$, $c_{1}$ and $c_{2}$ are the fitting parameters. The second term represents the Lorentzian transmission spectrum far away from the EP, while the third term is the squared-Lorentzian contribution near the EP~\cite{Kenta2021}. The difference  between the two eigenfrequencies is determined as $\Delta E_{\text{x}} = \text{Re}(E_{+,\text{x}}) - \text{Re}(E_{-,\text{x}})= \Delta_{+,\text{x}} - \Delta_{-,\text{x}}$. A eigenfrequency splitting manifests the $\mathcal{PT}$-symmetric phase transition~\cite{Bender2007}. 

Figures ~\ref{fig:FIG3}(a) and (b) show the numerical results and fittings of the transmission spectra at $g_\text{CCW} = 15.09$ and $g_\text{CW} = 2.25$ in the forward and backward cases, respectively, indicated by the dashed vertical line in Fig.~\ref{fig:FIG3}(c). The fittings with Eq.~\ref{eq:Eq5} are in excellent agreement with simulation results. The backward transmission splits and gives $\Delta_{+,\text{CW}} -\Delta_{-,\text{CW}} = 1.84$, close to the theoretical value $2.02$ according to the single-excitation model Eq.~\ref{eq:Eq3}. It means that the backward subsystem is in the unbroken $\mathcal{PT}$-symmetric phase. In contrast, we obtain $\Delta_{+,\text{CCW}} -\Delta_{-,\text{CCW}} = 0$ in the forward case. Thus, the eigenfrequencies are degenerate, manifesting the broken $\mathcal{PT}$ symmetry. 

The eigenfrequencies can be extracted with high precision from the transmission spectra with Eq.~\ref{eq:Eq5}.
Figure~\ref{fig:FIG3}(c) displays the eigenfrequencies as a function of the coupling strengths in the forward and backward cases. The numerical values are in good coincidence with the single-excitation model. Phase transitions occurs at EPs. Interestingly, the cQED system shows nonreciprocal $\mathcal{PT}$-symmetric phase transition as the atom-resonator interaction increases. When the interaction is weak that $g_\text{CW} < g_\text{EP, CW}$, the system is in the broken $\mathcal{PT}$-symmetric phase for the CW and CCW cases. The symmetry maintains in both cases when the interaction is strong enough that $g_\text{CCW} > g_\text{EP, CCW}$. Remarkably, the system exhibits nonreciprocal $\mathcal{PT}$-symmetric phase transitions at $g_\text{EP, CW}$ and $g_\text{EP, CCW}$. At $g_\text{CW} = g_\text{EP, CW}$, phase transition happens in the backward subsystem, but the $\mathcal{PT}$ symmetry remains in the forward one. As the atom-resonator coupling increases to $g_\text{CCW} = g_\text{EP, CCW}$,  the $\mathcal{PT}$-symmetric phase transition occurs in the forward subsystem. In the middle that $g_\text{EP, CW} < g_\text{CW} (g_\text{CCW}/\sqrt{45}) <g_\text{EP, CCW}/\sqrt{45}$, the forward and backward subsystems lie in essentially different phases, showing nonreciprocity. The former one is in a phase with broken $\mathcal{PT}$ symmetry, while the latter remains the $\mathcal{PT}$ symmetry.

\begin{figure}
  \centering
  \includegraphics[width=1.0\linewidth]{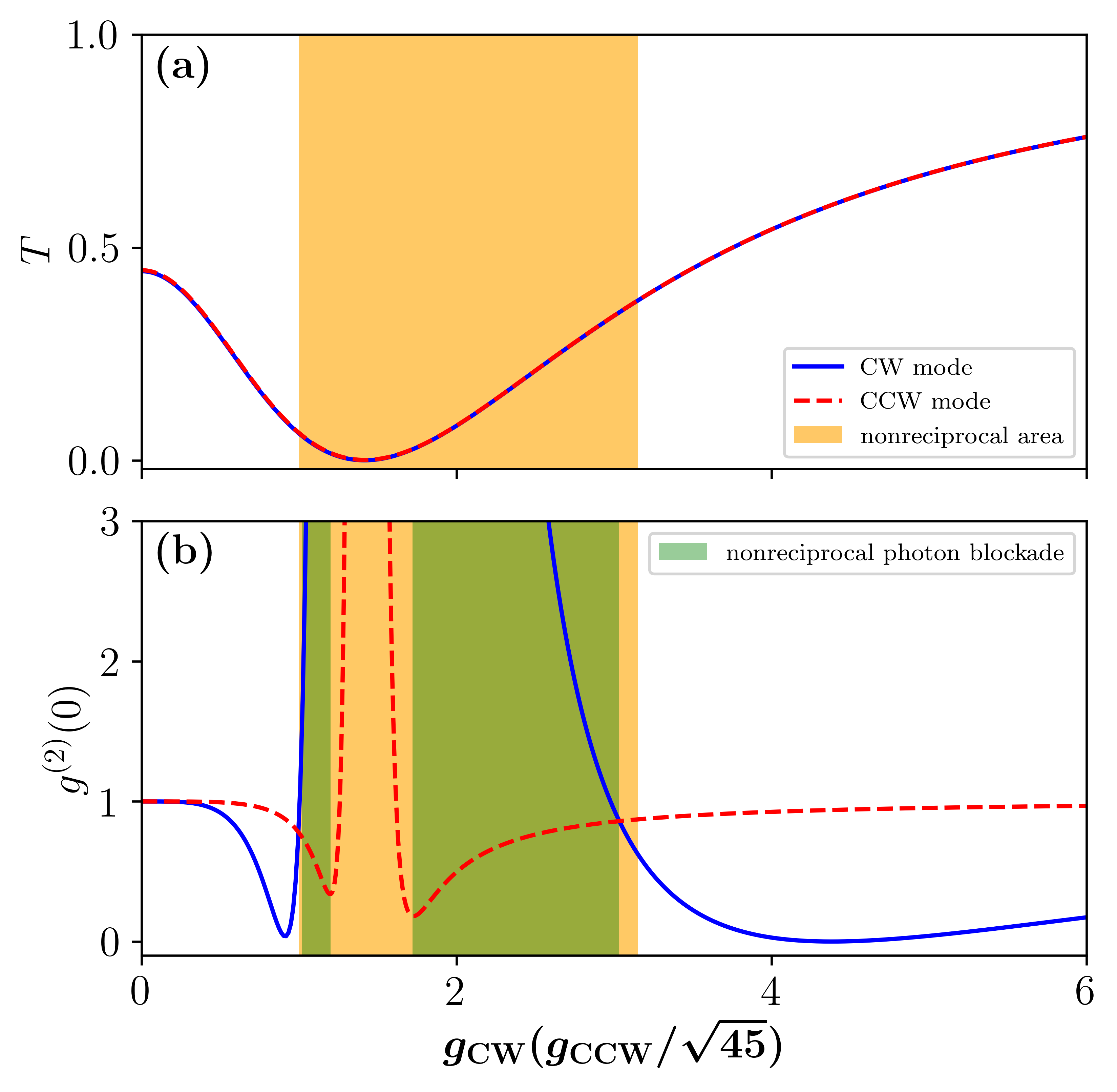} \\
\caption{(a) The resonator transmission with $\Delta = 0$ with respect to the atom-resonator coupling strength under the CW (blue line) and the CCW (red dashed line) mode excitations. Orange shaded area represents nonreciprocal $\mathcal{PT}$-symmetric phase transition area. (b) The second-order quantum correlation values of the system versus the coupling strength under the CW and the CCW mode excitations. Green shaded area represents nonreciprocal photon blockade.  }
\label{fig:FIG4}
\end{figure}

\emph{Nonreciprocal photon blockade}.---Nonreciprocal $\mathcal{PT}$-symmetric phase transitions are available when the coupling or the dissipation are different in the CW and CCW cases. By combining the chiral coupling and dissipation, we obtain nonreciprocal photon blockade in the nonreciprocal phase region.
As shown in Fig.~\ref{fig:FIG4}(a), the resonant transmission ($\Delta_p =0$) are equal for opposite inputs as the coupling strengths varies. The system transmission is reciprocal.
In Fig.~\ref{fig:FIG4}(b), we study the quantum correlation properties of the system through numerical simulations. We calculate the equal-time second-order correlation function $g^{(2)}_\text{x}(0) = \langle a_\text{out,x}^{\dagger}a_\text{out,x}^{\dagger}a_\text{out,x}a_\text{out,x}\rangle/\langle a_\text{out,x}^{\dagger}a_\text{out,x}\rangle^2$ with $\text{x} \in \{\text{CW, CCW}\}$, where $a_\text{out,CW} = \alpha_\text{in} - \sqrt{2\kappa_\text{ex}}a_\text{CW}$ and $a_\text{out,CCW} = \beta_\text{in} - \sqrt{2\kappa_\text{ex}}a_\text{CCW}$~\cite{Barak2008}. The system shows nonreciprocal photon blockade in two regions. When $1.02< g_\text{CW}<1.20$ or $1.72< g_\text{CW}<3.03$, photon blockade is attained in the forward case, manifesting by $g^{(2)}_\text{CCW}(0)<1$. By contrast, the backward subsystem exhibits strong photon bunching, indicated by $g^{(2)}_\text{CW}(0) > 1$. Thus, our system exhibits a pure nonreciprocal quantum behavior within the nonreciprocal quantum phase transition region.

%The required atom-resonator system can be made 

\emph{Conclusion and discussion}.---We have theoretically proposed a non-Hermitian cQED system for achieving nonreciprocal quantum phase transition. The underlying mechanism is either the chiral coupling or the chiral dissipation induced by the broken $\mathcal{T}$ symmetry due to the spin-momentum locking in a photonic microstructure. Spin-momentum locking has been the basis of numerous exotic fundamental physics such as photonic Hall effect~\cite{Konstantin2015}. In combination with chiral light-matter interaction, it has also become a working horse for constructing chiral quantum optical systems and realizing optical isolation.  It is yet to be explored for nonreciprocal phase transition. Unlike the chiral interaction, little attention is paid to the chiral dissipation. Here, nonreciprocal quantum phase transition and photon blockade are unveiled by exploring spin-momentum locking and the chiral dissipation. The discovered nonreciprocal photon blockade without using a spinning part may pave the way for unconventional quantum information processing in an integrated platform. 

%Acknowledgement
This work was supported by National Key R\&D Program of China (Grant No.~2019YFA0308700), Innovation Program for Quantum Science and Technology (Grant No.~2021ZD0301400), the National Natural Science Foundation of China (Grants No.~11890704, No.~92365107, and No.~12305020), the Program for Innovative Talents and Teams in Jiangsu (Grant
No.~JSSCTD202138), the Shccig-Qinling Program, the China Postdoctoral Science Foundation (Grant
No.~2023M731613), and the Jiangsu Funding Program for Excellent Postdoctoral Talent
(Grant No.~2023ZB708).

% \bibliography{Reference}

%apsrev4-2.bst 2019-01-14 (MD) hand-edited version of apsrev4-1.bst
%Control: key (0)
%Control: author (8) initials jnrlst
%Control: editor formatted (1) identically to author
%Control: production of article title (0) allowed
%Control: page (0) single
%Control: year (1) truncated
%Control: production of eprint (0) enabled
\providecommand{\noopsort}[1]{}\providecommand{\singleletter}[1]{#1}%

\end{document}